\documentclass[12pt,preprint]{emulateapj}
\usepackage{amsmath}
\usepackage{amssymb}
\usepackage{graphicx}

\slugcomment{Submitted V2 - December 16th, 2009}
\lefthead{Y.-F. Jiang et al.} \righthead{Synchrotron Emission from
Elliptical Galaxies Consequent to AGN Outbursts}

\begin{document}
\title{Synchrotron Emission from Elliptical Galaxies Consequent to AGN Outbursts}

\author{Yan-Fei Jiang\altaffilmark{1}, Luca Ciotti\altaffilmark{2},
Jeremiah P. Ostriker\altaffilmark{1,3} and Anatoly
Spitkovsky\altaffilmark{1} } \affil{$^1$Department of Astrophysical
Sciences, Princeton University, NJ, USA, 08544}
\affil{$^2$Department of Astronomy, University of Bologna, via
Ranzani 1, I-40127, Bologna, Italy} \affil{$^3$IoA, Cambridge, UK}

\begin{abstract}
Both radiative and mechanical feedback from Active Galactic Nuclei
have been found to be important for the evolution of elliptical
galaxies. We compute how a shock may be driven from a central black
hole into the gaseous envelope of an elliptical galaxy by such
feedback (in the form of nuclear winds) using high resolution 1-D
hydrodynamical simulations. We calculate the synchrotron emission
from the electron cosmic rays accelerated by the shocks (not the
jets) in the central part of elliptical galaxies, and we also study
the synchrotron spectrum's evolution using the standard diffusive
shock acceleration mechanism, which is routinely applied to
supernova remnants. We find quantitative consistency between the
synchrotron radio emission produced via this mechanism with extant
observations of elliptical galaxies which are undergoing outbursts.
Additionally, we also find that synchrotron optical and X-ray
emission can co-exist inside elliptical galaxies during a specific
evolutionary phase subsequent to central outbursts. In fact, our
calculations predict a peak synchrotron luminosity of $\sim
1.3\times 10^6\ L_{\odot}$ at the frequency $5$ GHz (radio band), of
$\sim 1.1\times 10^6\ L_{\odot}$ at $4.3\times10^{14}$ Hz
(corresponding to the absolute magnitude -10.4 in R band), and of
$\sim 1.5\times 10^{7}\ L_{\odot}$ at $2.4\times10^{17}$ Hz (soft
X-ray, $0.5$ --- $2.0$ keV band).
\end{abstract}


\section{Introduction}
Is there direct evidence for feedback from Active Galactic Nuclei
(hereafter AGN) available from the non-thermal emission of the
central regions of massive galaxies? It has been observed that 10-20
percent of AGN are radio-loud and that the host galaxies of
radio-loud AGNs are usually massive elliptical galaxies (e.g.,
\citealt{Hooperetal1995}; \citealt{Best2007}). The radio emission in
radio galaxies and quasars comes from a variety of morphological
structures, with different characteristic sizes. For example, there
are compact radio cores (usually with flat power law spectra)
coincident with the nucleus of the associated optical objects, as
well as large scale radio lobes. Other systems show radio lobes
which may extend to several hundred kiloparsecs to a few megaparsecs
and usually show steep power law radio spectra (e.g.,
\citealt{Kellermannetal1994}; \citealt{krolik1999};
\citealt{kembhavinarlikar1999}; \citealt{Gilbertetal2004};
\citealt{Mullinetal2006}). The typical core radio luminosity for
FR-II sources at 5GHz is $\sim 10^{30}$ erg s$^{-1}$ Hz$^{-1}$,
while the total luminosity at 178 MHz is $\sim 10^{33}$ erg s$^{-1}$
Hz$^{-1}$ (e.g., \citealt{Chiabergeetal2000}), with the spectral
index (absolute value) typically smaller than $0.5$ at the very
center and $\sim 0.7-0.8$ at a larger radius. In addition to the
radio observations, there are also observations in other bands for
radio galaxies. For example, in the R-band 3CR Snapshot Survey of
252 radio galaxies, the wind-like structures seen in the contour
figures of \cite{marteletal1999} may be interpreted as produced by
nuclear winds.
 Kiloparsec-scale winds are also observed in high redshift
quasars (e.g., \citealt{dekooletal2001}; \citealt{Nesvadba2009};
\citealt{Nesvadbaetal2008}; \citealt{moeetal2009};
\citealt{Cattaneoetal2009} and references therein). Optical
synchrotron jets are also confirmed in many sources (e.g.,
\citealt{marteletal1999}) and can extend to distances far outside of
the body of the galaxies. Furthermore, ultraviolet and X-ray
observations have also found high-velocity outflows in some quasars
(e.g., \citealt{Nesvadbaetal2008}; \citealt{Hamannetal2008};
\citealt{Nesvadba2009}; \citealt{Alexanderetal2009};
\citealt{Chartasetal2009}). Finally, although the evidence is not
absolutely conclusive, there are observations (e.g.,
\citealt{Chiabergeetal2000}) indicating optical synchrotron from the
cores of some active elliptical galaxies. The obvious question is
``what processes accelerate cosmic rays (hereafter CR) electrons up
to the energies required to produce the observed synchrotron
emission?"

Synchrotron emission is believed to be the mechanism responsible for
the radio emission in AGNs, based on their strong polarization
(e.g., \citealt{krolik1999}). The relativistic electrons, which are
required for the synchrotron radio emission, can be accelerated by
several sources. Jets are observed in most of the radio-loud AGNs
and most investigators believe that those jets produce this radio
synchrotron emission (e.g., \citealt{Rees1971};
\citealt{Blandfordrees1974}; \citealt{SironiSocrates2009}; also see
the review \citealt{MirabelRodriguez1999} and references therein).
However, the energy carried by the jets is only dissipated in small
areas near the end point or within internal reflected shocks. As the
jets are driven outwards with relativistic velocity, the centers of
the radio contours from the jets do not overlap with the centers of
the galaxies (see, e.g., \citealt{Kellermannetal1994};
\citealt{Mullinetal2006}). So, although jets are very likely
responsible for the displaced radio lobes, there must be other
mechanisms responsible for the observed radio core emission at the
centers of FR-II galaxies. Furthermore, in order to explain
synchrotron emission in galaxies without jets, other sources able to
accelerate electrons are needed. In fact, jets are observed in
$\sim$ 80\% of FR-I radio galaxies, in $\sim$ 40-70\% radio quasars,
and even fewer in highly luminous FR-II galaxies (e.g.,
\citealt{Bridleperley1984}; \citealt{kembhavinarlikar1999}).

This paper addresses the inevitable generation of radio synchrotron
sources in elliptical galaxies consequent to central outbursts. In
fact, it is now widely accepted that supermassive black holes
(SMBHs) within a mass range of $10^6$ $\sim$ $10^{9.5}$ M$_{\odot}$
reside at the center of bulges and elliptical galaxies, and that
feedback from these SMHBs can profoundly affect the formation and
evolution of these galaxies (e.g., \citealt{KormendyRichstone1995};
\citealt{Silk Rees1998}; \citealt{Fabian1999};
\citealt{BurkertSilk2001}; \citealt{King2003};
\citealt{Granatoetal2004}; \citealt{Springeletal2005};
\citealt{Sazonovetal2005}; \citealt{Hopkinsetal2006};
\citealt{Huetal2006}). This is strongly supported by the observed
relationships between the masses of the SMBHs and various properties
of their host galaxies (e.g., \citealt{Magorrianetal1998};
\citealt{Gebhardtetal2000}; \citealt{Ferraresemerritt2000};
\citealt{Tremaineetal2002}; \citealt{McLureDunlop2002};
\citealt{GrahamDriver2007}; \citealt{Laueretal2007};
\citealt{loujiang2008}). In particular, in a series of works based
on high resolution numerical simulations which include both
radiative and mechanical feedback (\citealt{CiottiOstriker1997},
2001, 2007; \citealt{Ciottietal2009}, hereafter Paper I), it has
been shown how the mass loss from evolving stars can drive
significant nuclear activity, characterized by strong and recurrent
nuclear bursts, even in the absence of merging. This model supports
the idea that the ``cooling flow" phases (e.g.,
\citealt{PetersonFabian2006}) and quasar phases are different
aspects of the evolution of a normal elliptical galaxy. In the
calculations, it is found that accretion occurs in bursts, during
which both radiative and mechanical output from the central SMBH
pushes matter out and drives shocks into the galactic gas.
 The outbursts typically combine three physically separate
phenomena: central star bursts, mechanical feedback from winds
emanating from broad-line regions surrounding the SMBH, and
radiative feedback from absorption and scattering of hard X-ray
photons.  The resulting shocks are similar in kind (but more modest
in degree) to the predicted outflowing blast waves produced by AGNs
after galaxy merging (e.g., see \citealt{DiMatteoetal2005};
\citealt{Springeletal2005}; \citealt{Naabetal2006};
\citealt{Sijackietal2007}; see also \citealt{Johanssonetal2009}).

In this paper, we focus attention on the observational properties of
the emitted synchrotron emission from the shocks that result from
the recurrent bursts.
 In fact, since the
computed outflow velocities (up to several thousand kilometers per
second) and densities (0.1 to 10 particles per cubic centimeter) are
similar to what is observed in Galactic SNRs such as Tycho and
Kepler (e.g., \citealt{Cassamchenaetal2007};
\citealt{Dickeletal1988}), one would expect that the same processes
 as those we see acting in
SNRs on the pc scale (specifically at the shocks' surfaces, Wang
2008) would act on the kpc scale in galaxies. One can expect there
to be efficient acceleration of both ionic and electronic cosmic
rays via the well-known diffusive shock acceleration mechanism (also
known as first-order Fermi shock acceleration, e.g.,
\citealt{BlandOstriker1978}; \citealt{BlandfordEichler1987};
\citealt{Berezhkoellison1999}; see also
\citealt{TreumannJaroschek2008}). Then, just as the accelerated CR
electrons produce synchrotron emission through standard interactions
with the (possibly amplified) magnetic field within normal SNRs, so
we should expect very similar processes to occur in the ISM of
elliptical galaxies after outbursts. To our knowledge, this has not
yet been studied in quantitative detail. Using
 diffusive shock acceleration theory, we can make a rough
estimate of the scaling relation for luminosity, because the
synchrotron luminosity only depends on the CR electron density, the
magnetic field $B$ and the shock velocity. If the density and
velocity of the shock are fixed, the luminosity is determined by the
volume occupied by CR electrons and the magnetic field. The extent
of emitting region near the shock is determined by the synchrotron
losses of the electrons, which are proportional to $B^2$, so that
the volume occupied by CR electrons goes with $r^2/B^2$ ($r$ is the
radius of the shock). Because power per volume radiated by the
electrons goes like $B^2$, the total luminosity scales approximately
as $r^2$. As the size of an elliptical galaxy is $\sim100$ times
larger than the SNRs, and the densities and shock velocities in the
cores of elliptical galaxies are expected to be very similar to
those in SNRs, we expect the synchrotron luminosity from elliptical
galaxy cores to be $\sim 10^4$ --- $10^5$ larger than the luminosity
from SNRs. Our detailed calculations (see \S 3) confirm these rough
scaling estimates.

In this paper, we calculate synchrotron emission from the
relativistic electrons accelerated by shocks during the evolution of
an elliptical galaxy and we also show how the synchrotron spectrum
evolves with time. The Ciotti \& Ostriker model for the evolution of
elliptical galaxies provides us with a specific scenario for
computing the shocks, which are formed repeatedly due to the
feedback of the central SMBHs. With the propagation of the shock
into the interstellar medium (ISM), electrons will be accelerated at
different positions (via the standard diffusive shock acceleration
mechanism), and high energy particles can be found at large radii.
This is very different from the concept sometimes proposed that CRs
are accelerated at the center and escape to large radius,
since such CR electrons would suffer from large adiabatic losses and
become relatively ineffective radiators. In \S 2 we present the
method to calculate the  accelerated electron spectrum and
associated synchrotron emission. In \S 3 we show the result from a
single shock taken from a numerical simulation. The main results are
summarized in \S 4 and discussions and observational results of our
model are given in \S 5.

\section{Calculation Method}
In this Section, we describe the numerical model adopted for the
simulation and we give the formulae for the diffusive shock
acceleration mechanism and  synchrotron emission used in this paper.

\subsection{The Hydrodynamical Model}
\label{codemodel}

The details of hydrodynamical simulations of radiative plus
mechanical feedback and how the kinetic energy, momentum and mass of
the wind are transferred to the ISM in combined (i.e., radiative
plus mechanical feedback) models are described in Paper I, while the
exploration of parameter space in combined models is postponed to
\cite{Ciottietal2009c} (hereafter Paper III). We recall that the
code is 1-D, and that, as in Paper I, a simplified version of
mechanical feedback is adopted\footnote{In practice, the time
derivative in equation (28) of Paper I is set to zero.}, which is
similar to that used by others (e.g., \citealt{DiMatteoetal2005})
studying AGN feedback. The galaxy model is a Jaffe stellar
distribution embedded in a dark halo so that the total density
profile decreases as $r^{-2}$ (\citealt{Ciottietal2009b} and
references therein). The cooling and heating functions
(\citealt{Sazonovetal2005}) include photoionization plus Compton and
line heating, while the SMBH accretion rate is mediated by a
circumnuclear accretion disk modelized at the level of subgrid
physics. Finally, the mass return rate from evloving stellar
populations is computed by using the detailed prescriptions of
stellar evolution. A discussion of the limitations of 1-D simulation
is deferred to \S \ref{discussion}.

\begin{figure}
\includegraphics[height=8cm]{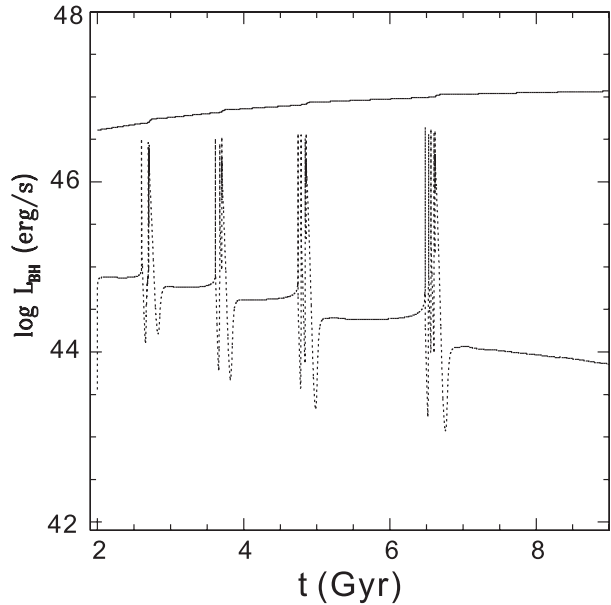}\\
\caption{Time evolution of the SMBH bolometric accretion luminosity
L$_{\rm{BH}}$ for model B3$_{02}$. The nearly horizontal solid line
is the Eddington luminosity L$_{\rm{Edd}}$. There are four major
bursts during the evolution history of the model, and each major
burst is composed of several sub-bursts. } \label{examtimeburst}
\end{figure}

In the following discussion, we focus on a specific model labeled
B3$_{02}$ (Paper I; Paper III, in preparation), which models the
radiation and mechanical feedback from AGNs to the host elliptical
galaxies. This model has an initial stellar mass
M$_{\star}$=$2.87\times10^{11}$ M$_{\odot}$ and an initial SMBH mass
M$_{\rm{BH}}$=$2.87\times10^{8}$ M$_{\odot}$. The galaxy effective
radius is R$_{\rm{e}}$=6.9 kpc and the aperture central velocity
dispersion is $\sigma_0$=260 km s$^{-1}$. The first grid point in
the code corresponds to $R_{\rm{min}}$ =5 pc.
The simulation starts at 2 Gyr and lasts to 14.5 Gyr. The time
evolution of the SMBH bolometric accretion luminosity is shown in
Fig. \ref{examtimeburst}.

The computations show that a cooling instability
(\citealt{Field1965}) starts as a cool ($T\sim 10^4 $ K) and dense
shell at $r\sim$ 1 kpc; this shell collapses, feeds gas into the
center, and the SMBH responds with a major burst. In more detail,
the infalling shell compresses the gas inside the collapsing volume,
so that more and more gas is accreted by the SMBH even before the
cold shell reaches the center. The luminosity of SMBH increases
rapidly with time, and when it reaches $\sim$1 \% of Eddington
luminosity, the infalling matter is pushed out by the radiation from
the SMBH (e.g., \citealt{Ostrikeretal1976};
\citealt{Milosavljevicetal2009}). At the same time, a wind appears
from the circumnuclear disk and a shock is driven into the ISM. The
shock hits the still infalling cold shell at a very high Mach
number, due to the low temperature of the shell. The shock continues
to propagate into the galaxy; gradually its energy is deposited into
the ISM, and it dies away with time. In the meantime, reverse shock
waves carry fresh gas to the center, and several sub-bursts are
generated. Finally, the last burst ends the sequence. The duty cycle
--- the fraction of time that the central SMBH is in the outburst mode
with L$_{\rm{BH}}$ $>$ L$_{\rm{Edd}}$/30
--- is of the order of $ \sim$ 5 \%, consistent with the fraction of
elliptical galaxies seen as quasars. As commonly found in combined
(i.e., radiative plus mechanical feedback) models discussed in full
extent in Paper III, the temporal structure of each major burst is
highly organized into a series of sub-bursts (as apparent from the
temporal structure of each of the 4 major bursts in Fig.
\ref{examtimeburst}), while purely mechanical feedback models
present much simpler bursts, as can be seen by comparison with Figs.
2, 3 and 4 in Paper I. In our case, each shock will accelerate
electrons and protons.

\begin{figure}
\includegraphics[angle=0.,scale=0.5]{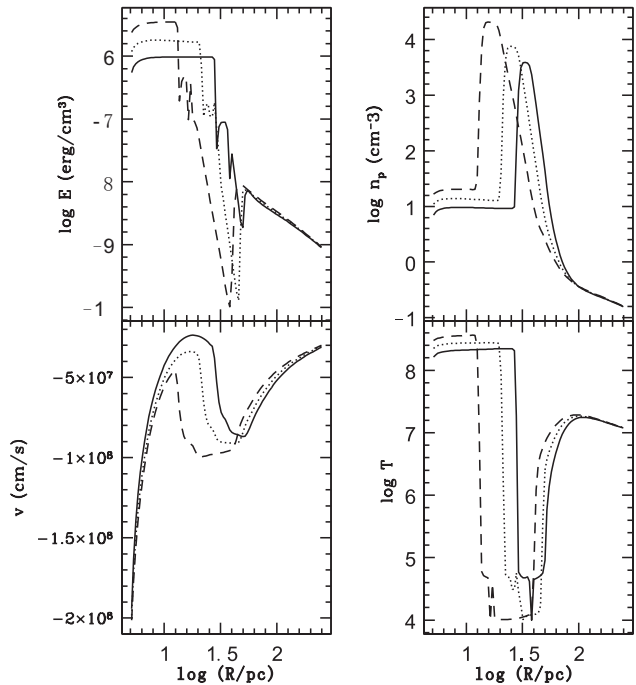}\\
\caption{Thermal instability and the falling cold shell: the radial
profile of the ISM energy per unit volume $E$, proton number density
$n_p$, radial velocity $v$, and temperature $T$. Solid, dotted and
dashed lines correspond to $\Delta t=0.204,\ 0.214$ and 0.224 Myr
after L$_{\rm{BH}}$ reached $0.01$ L$_{\rm{Edd}}$ ($6.48$ Gyr, see
Fig. \ref{examtimeburst}). This time is defined to be the zero
point. Note how the gas density in the central regions increases due
to compression before the cold shell (40 pc from the center at
$\Delta t=0.204$ Myr) reaches the center. }\label{examradi}
\end{figure}

\begin{figure}
\includegraphics[angle=0.,scale=0.5]{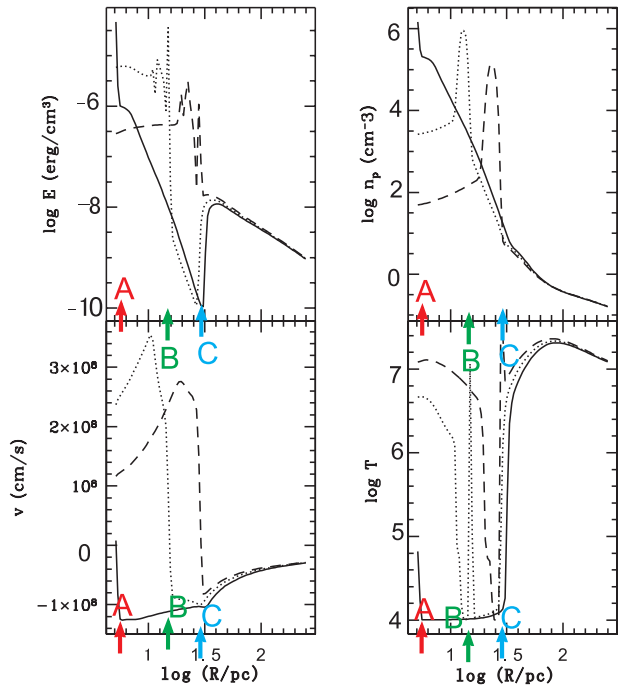}\\
\caption{Post outburst: the radial profile of the physical variables
when a burst appears at $\Delta t_1=0.235$ Myr. Solid, dotted and
dashed lines correspond to $\Delta t=0.235,\ 0.239$ and 0.243 Myr.
The fluid radial velocity changes direction from inward to outward
between $\Delta t_1=0.235$ Myr to $\Delta t_2=0.239$ Myr. The cold
shell is still visible in the solid line at the center. The arrows
(A, B, C) indicate the positions of the shock at the three defined
moments (see Fig. \ref{divshock}); the shock initially accelerates
(A$\to$B) due to the SMBH driving from below and then decelerates
(B$\to$C) as adiabatic losses overcome the SMBH input.
}\label{examradiburst}
\end{figure}

Now we focus on the first sub-burst in the last major burst around
6.5 Gyr. The SMBH bolometric accretion luminosity L$_{\rm{BH}}$
reaches 1 \% of Eddington luminosity L$_{\rm{Edd}}$ at the time 6.48
Gyr,
and we define this time as the zero time point. With this time
origin,  L$_{\rm{BH}}$ reaches $0.1$ L$_{\rm{Edd}}$ around $0.234$
Myr later, and it peaks at 0.5L$_{\rm{Edd}}$ at $\Delta t=0.252$
Myr. Fig. \ref{examradi} shows the radial profile of the physical
variables at $\Delta t=0.204$ Myr, just before the peak of the
burst. We can see clearly the cold shell initially at $40$ pc
falling to the center. Fig. \ref{examradiburst} shows the appearance
of outflow with the solid line ($\Delta t=0.235$ Myr) still before
the peak of the burst: the wind pushed out by radiation from the
SMBH reaches 5.1 pc, with a velocity of only $100$ km/s (arrow A).
By the time the shock reaches $15$ pc (arrow B) the maximum velocity
has increased up to $3600$ km/s, as it is being steadily accelerated
from below, and it has hit the infalling cold shell, which results
in a very high Mach number shock.

\subsection{Acceleration of Particles at the Shocks}

Synchrotron emission from shocks based on the diffusive acceleration
mechanism has been applied to astrophysical systems on different
scales. In particular, for SNRs
there exist different nonlinear models for the structure and
evolution of cosmic ray modified (strong) shocks (e.g., see
\citealt{Koyamaetal1995}; \citealt{Berezhkoellison1999};
\citealt{Ellisonetal2000}; \citealt{Lazendicetal2004};
\citealt{Ellisonetal2004}; \citealt{BerezhkoVolk2006};
\citealt{Ellisonetal2007}; see the reviews by
\citealt{Reynolds2008}, \citealt{TreumannJaroschek2008}). With this
method, the synchrotron spectra at different radii and their
evolution with time can be computed and then compared with the
observations of SNRs, showing good agreement (e.g.,
\citealt{Reynolds1998}; \citealt{Lazendicetal2004};
\citealt{Cassamchenaetal2007}). Some researchers have also carried
out simulations on the scale of galaxy clusters to study the radio
haloes and relics based on the diffusive shock acceleration
mechanism in the limit of test particle theory (e.g.,
\citealt{Ensslinetal2007}; \citealt{Pfrommeretal2008}). Here, we
apply the same techniques to the case of wind-driven shocks in
elliptical galaxies.

Our CR shock acceleration scheme is based on well established
physical principles (the first order Fermi mechanism) but utilizes a
new, quite efficient numerical implementation, which allows us to
perform the analysis during post-processing instead of during the
runs, saving considerable computational time. We use the test
particle theory to calculate the electron spectrum at the shock,
which gives the basic characteristics of the non-thermal electrons
(e.g., \citealt{BlandOstriker1978}; \citealt{BlandfordEichler1987};
see also the recent review by \citealt{Reynolds2008}). Consistent
with test particle theory, the CRs are assumed to not change the
structure of the shocks, so that we can use the output of the
hydrodynamical code to calculate the synchrotron spectrum.

In order to compute the acceleration of the electrons and protons,
first we have to identify and follow the shocks as they move through
the ISM. Each shock is driven by the combined effect of radiation
and nuclear winds following a burst, and it propagates into the
ambient gas of the elliptical galaxy.  For each hydrodynamical
output file, we calculate the dimensionless compression factor
\begin{eqnarray}
\delta\equiv\frac{r}{|v|}\rm{div}(\mathbf{v})\
\label{shockcompression}
\end{eqnarray}
at every point on the grid. Here $r$ is the radius, $\mathbf{v}$ is
the fluid velocity vector and $|v_r|$ is the absolute value of
velocity\footnote{Only the radial component is non-zero due to
adopted spherical symmetry.}. As the shock strongly compresses the
gas, a local deep minimum of $\delta$ marks the position of the
shock. The upstream sonic Mach number of the shock is given by
\begin{eqnarray}
\mathcal{M}=\frac{|v-u_s|}{c_s}\ ,
\end{eqnarray}
where $u_s$ is the shock velocity relative to the galaxy and $c_s$
is the upstream sound speed. The shock velocity $u_s$ is numerically
estimated by determining the change in the shock position within a
time step. The upstream sound speed is calculated from the upstream
temperature $T$ and density $\rho$ as
\begin{eqnarray}
c_s=\sqrt{\left(\frac{\partial{P}}{\partial{\rho}}\right)_S}=\sqrt{\frac{\gamma
k_BT}{\mu m_p}}\ ,
\end{eqnarray}
where $P$ is the pressure, $k_B$ is the Boltzmann constant,
$\mu\cong 0.62$ is the mean molecular weight, $m_p$ is the proton
mass, and $\gamma=5/3$ is the adiabatic index of the gas.

As the magnetic field $B$, necessary for the synchrotron emission,
is not included in our hydrodynamical code, we must assume a
fiducial value based on physical arguments. In particular, for
elliptical galaxies, there is evidence that the ratio between the
non-thermal and thermal pressures is about 0.1---0.2 (e.g.,
\citealt{Churazovetal2008}; Churazov et al. in preparation, private
communication). Here we just assume the ratio between the thermal
and magnetic pressure to be
\begin{eqnarray}
\beta\equiv\frac{P}{B^2/(8\pi)}\cong10, \label{beta}
\end{eqnarray}
so that we can estimate the postshock magnetic field from the
thermal pressure.

Now we summarize the basic results of the test particle theory. The
1-D momentum spectrum of the non-thermal electrons is taken to be a
power law :
\begin{eqnarray}
f(p)=4\pi p^2\times \mathcal{C}_0\ p^{-q}=4\pi \mathcal{C}_0\
p^{2-q}, \label{momentumspec}
\end{eqnarray}
where $p$ is the momentum of the electrons and $\mathcal{C}_0$ is
the normalization of the spectrum, which will be determined later.
The exponent $q$ is given by\footnote{Here we neglect the Alfven
speed, with which the scattering centers presumably move, in
comparison to the flow speed. This correction is only important for
low Mach number shocks. Because synchrotron emission mainly comes
from strong shocks, neglecting Alfven speed will not change the
spectrum significantly (e.g., \citealt{capriolietal2009}).}
\begin{eqnarray}
q=\frac{3\tau}{\tau-1}\ ,
\end{eqnarray}
where the compression ratio $\tau$ is related to the Mach number
$\mathcal{M}$ as
\begin{eqnarray}
\frac{1}{\tau}=\frac{\gamma-1}{\gamma+1}+\frac{2}{\gamma+1}\frac{1}{\mathcal{M}^2}\
.
\end{eqnarray}
From the momentum distribution (\ref{momentumspec}), and from the
energy-momentum relation
\begin{eqnarray}
p=m_ec\times\sqrt{\Big(1+\frac{E}{m_ec^2}\Big)^2-1},
\end{eqnarray}
where $E$ and $m_e$ are the kinetic energy and the rest mass of
electrons respectively, and $c$ is the speed of light, we can
express the energy spectrum of the non-thermal electrons as
\begin{eqnarray}
N_0(E)=4\pi \mathcal{C}_0\ p^{2-q} \frac{dp}{dE}\ .
\label{energyspec}
\end{eqnarray}

The energy of non-thermal electrons ranges between $E_{min}$ (with
the corresponding minimum momentum $p_{min}$) and $E_{max}$ (with
maximum momentum $p_{max}$). Following what is done in the
computations of SNRs and galaxy clusters (e.g.,
\citealt{Berezhkoellison1999}; \citealt{Ensslinetal2007}), the
minimum momentum $p_{min}$ is related to the downstream temperature
$T_2$ by the injection parameter $x_{inj}$:
\begin{eqnarray}
p_{min}=x_{inj}\ \sqrt{2\times m_ek_BT_2} \ \ \ (\rm{cgs}).
\label{minmomentum}
\end{eqnarray}
For electrons, the maximum energy at the shock is determined by the
balance between radiative loss and acceleration gain, which
gives\footnote{In a typical elliptical galaxy, the energy density of
IR background and star light is usually small compared with the
downstream magnetic energy density. So we ignore the electrons'
inverse Compton cooling here.} (e.g., \citealt{Webbetal1984};
\citealt{Reynolds2008}):
\begin{eqnarray}
E_{max}&=&71.65\ (u_s/10^8)\times(B_2/10^{-6})^{-0.5}\ \rm{erg}\ (\rm{cgs})\nonumber\\
&=&44.72\ u_{s,10^3\rm{km}/\rm{s}}\times (B_{2,\rm{\mu G}})^{-0.5}\
\rm{TeV}.
\end{eqnarray}
where $B_2$ is the downstream magnetic field. Actually, in the test
particle theory, the synchrotron spectrum is not very sensitive to
the maximum energy $E_{max}$. But the cutoff of the power-law
spectrum does depend on $E_{max}$. In order to determine the
normalization of the electron spectrum, we follow the approach
adopted in simulations of galaxy clusters (e.g.,
\citealt{Ensslinetal2007}; \citealt{Pfrommeretal2008}) and smoothly
connect the power-law momentum spectrum of electrons to the
downstream thermal spectrum at the momentum $p_{min}$.
Then the coefficient $\mathcal{C}_0$ in equation
(\ref{momentumspec}) is given by
\begin{eqnarray}
\mathcal{C}_0=\frac{n_{e,2}}{(2\pi m_ek_B T_2)^{3/2}}\ p_{min}^q\
e^{-p_{min}^2/(2m_ek_BT_2)}\ , \label{coef0}
\end{eqnarray}
where $n_{e,2}$ is the downstream electron number density.

The test particle result is valid only when the energy density of
the non-thermal particles is small enough compared to the thermal
energy density so that the CRs do not significantly change the
structure of the shock. However, for a strong shock with a very high
Mach number (e.g., when the shock encounters the cold shell),
calculation with the above method will easily produce an electron
energy density comparable to the downstream thermal energy. So in
our calculations we have to choose appropriate normalization of the
electron momentum spectrum for different Mach numbers, and we adopt
the method used in \cite{Ensslinetal2007} and
\cite{Pfrommeretal2008}. First, we obtain the electron energy
spectrum from equations (\ref{energyspec}) and (\ref{coef0}), so
that the energy density of the non-thermal electrons $K_{e,0}$ is
\begin{eqnarray}
K_{e,0}=\int_{E_{min}}^{E_{max}}\ EN_0(E)dE\ . \label{elecenergy0}
\end{eqnarray}
Then, the ratio between the non-thermal electron energy density and
the downstream thermal energy density (in the test particle theory
approach) is
\begin{eqnarray}
\xi_{lin}=\frac{K_{e,0}}{1.5n_2k_BT_2}\ , \label{xilin}
\end{eqnarray}
where $n_2$ is the downstream total number density. The ratio
$\xi_{lin}$ can be very large for a shock with large $\mathcal{M}$.
Now we set a maximum ratio between the non-thermal electron energy
density and the downstream thermal energy density to
$\xi_{max}=0.05$ (a value suggested by simulation of
\citealt{Keshetetal2003}; also see \citealt{Pfrommeretal2008}) and
define a parameter $\chi\equiv\xi_{lin}/\xi_{max}$. The modified
normalization of the electron momentum spectrum we use in our
calculation becomes
\begin{eqnarray}
\mathcal{C}=(1-e^{-\chi})\chi^{-1}\ \mathcal{C}_0\ ,
\label{saturation}
\end{eqnarray}
so that $\mathcal{C}\rightarrow \mathcal{C}_0$ for $\chi\rightarrow
0$, while $\mathcal{C}\rightarrow\mathcal{C}_0/\chi$ for very large
values of $\chi$. In other words, if $\xi_{lin}$ is very small
compared with $\xi_{max}$ ($\chi\ll 1$), then
$\mathcal{C}\sim\mathcal{C}_0$ and the energy density ratio is
approximately $\xi_{lin}$. However, if $\xi_{lin}$ is very large
compared with $\xi_{max}$ ($\chi\gg 1$), the final energy density
ratio between the non-thermal electrons and the downstream thermal
energy density cannot exceed $\xi_{max}$. Roughly speaking, the
modification is important when $\mathcal{M}$ is larger than $10$.
Therefore, equation (\ref{energyspec}) becomes
\begin{eqnarray}
N(E)=4\pi \mathcal{C}\ p^{2-q} \frac{dp}{dE}\ ,\label{realspec}
\end{eqnarray}
and the energy density and number density of the non-thermal
electrons are given by
\begin{eqnarray}
K_{e}=\int_{E_{min}}^{E_{max}}EN(E)dE\
,N_{e}=\int_{E_{min}}^{E_{max}}N(E)dE. \label{elecenergy}
\end{eqnarray}
The finally ratio between the non-thermal electron energy density
and downstream thermal energy density $\xi$ is the same as equation
(\ref{xilin}) with $K_{e,0}$ replaced by $K_e$. In particular, the
injection coefficient $\eta$, a parameter determining the
acceleration efficiency modeled in the diffusive shock acceleration
mechanism (e.g., \citealt{Ensslinetal2007}), is defined here to be
\begin{eqnarray}
\eta\equiv\frac{N_{e}}{n_{e,2}}\ . \label{eta}
\end{eqnarray}
Another thing which must be emphasized is that here we do not give
the energy density of the non-thermal protons, which is unnecessary
for our purpose. By using test-particle theory we are assuming
non-modified shocks, and the CR energy density is less than 10\% of
thermal energy density by assumption\footnote{Some observations show
that  in SNRs 10\% of the mechanical energy of the explosion can be
converted into CRs (e.g., \citealt{Aharonianetal2004}).}.
This is only a first order approximation and a first step to
determine
whether a more advanced model is merited. 

\subsection{The Spectrum after the Shocks} 

The test particle framework described above gives the energy
spectrum of the non-thermal electrons at the shock, and we need to
compute the evolution of the spectrum further downstream. For
simplicity, we assume no diffusion in our calculation. This means
that once the accelerated electrons are generated at the shock, they
are frozen
to the fluid element. So, 
while the shock propagates outwards, the non-thermal electrons are
left behind and move with the fluid. These non-thermal electrons
suffer radiative and adiabatic losses, and the electron energy
spectrum will change with time. The changes of the electron energy
and number density are well-known (e.g.,
\citealt{Reynolds1998}; \citealt{Cassamchenaetal2007}). 
We will describe our numerical implementation of the spectral
evolution formulae of \cite{Reynolds1998}. We assume that an
electron with energy $E_0$ is generated at time $t_0$ at radius
$r_0$, where the mass density is $\rho_0$. At a later time $t$, the
electron moves to the new position $r$, where the mass density is
$\rho_t=\rho(r,t)$, so that the compression ratio of the fluid
element is
\begin{eqnarray}
\alpha(r,t)\equiv\frac{\rho_t}{\rho_0}\ .
\end{eqnarray}
After the losses, the energy of the electron at time $t$ and radius
$r$ is
\begin{eqnarray}
E(r,t)=\alpha^{1/3}\frac{E_0}{1+\Theta E_0}\ , \label{changenergy}
\end{eqnarray}
where the quantity $\Theta$ is the result of the following integral
\begin{eqnarray}
\Theta(t)\equiv\mathcal{A}\int_{t_0}^t
B_{\rm{eff}}^2(t)\alpha^{1/3}(t)dt\ . \label{Theta}
\end{eqnarray}
The coefficient in front of the integral is $\mathcal{A}
\equiv4e^4/(9m_e^4c^7)=1.57\times10^{-3}\ (\rm{cgs})$ ($e$ is the
electron's charge ) and
\begin{eqnarray}
B_{\rm{eff}}\equiv\sqrt{B^2+B^2_{\rm{cbr}}(1+z)^4}\ ,
\end{eqnarray}
where $B_{\rm{cbr}}\equiv3.27\ \mu G$ is the magnetic field strength
with energy density equal to that of the CMB at the redshift $z=0$.

Each of the non-thermal electrons generated by the shock will change
its initial energy according to the above equations, and the energy
spectrum of these electrons when they are generated is given by the
test particle result (equation \ref{realspec}). As the number of
electrons is conserved, at time $t$ and radius $r$ the energy
spectrum of these electrons becomes
\begin{eqnarray}
N(E)_{r,t}=N(E_0)\alpha^{-2/3}\Big[\frac{E_0(E)}{E}\Big]^2\ .
\label{timespec}
\end{eqnarray}
Note that $E_0$ as a function of $E$ is determined by inverting
equation (\ref{changenergy}). In particular, initially the maximum
energy and minimum energy of the electrons are $E_{max,0}$ and
$E_{min,0}$. Subsequently, the maximum energy and minimum energy
will also change to the new maximum energy $E_{max}$ and minimum
energy $E_{min}$ according to equation (\ref{changenergy}).

Once we obtain the energy spectrum of the electrons, we can
calculate the resulting synchrotron spectrum, i.e., the emissivity
of the population of electrons in the energy interval $E_{min}\le
E\le E_{max}$
\begin{eqnarray}
J_{\nu}=\int_{E_{min}}^{E_{max}}\ \mathcal{P}(\nu)\ N(E)dE\ \ \
\hbox{erg}\ \hbox{s}^{-1} \hbox{cm}^{-3} \hbox{Hz}^{-1}.
\label{totalpower}
\end{eqnarray}
Then the luminosity $L_{\nu}$ can be calculated by integration of
$J_{\nu}$ over the whole volume. In the above equation, the
synchrotron emission from a single electron is given by
\begin{eqnarray}
\mathcal{P}(\nu)=\frac{\sqrt{3}e^3B_{\bot}}{m_ec^2}\frac{\nu}{\nu_c}\int_{\nu/\nu_c}^{\infty}K_{5/3}(x)dx\
\hbox{erg}\ \hbox{s}^{-1}\hbox{Hz}^{-1}, \label{synchro1}
\end{eqnarray}
where $K_{5/3}(x)$ is a modified Bessel function of the second kind,
and
\begin{eqnarray}
\nu_c=\frac{3eB_{\bot}}{4\pi m_e^3c^5}E^2
\end{eqnarray}
is the critical frequency, where $E$ is the energy of the electron
and $B_{\bot}$ is the magnetic field component perpendicular to the
line of sight (e.g., \citealt{BlumenthalGould1970};
\citealt{Lazendicetal2004}). In our calculation $B_{\bot}$ is
approximated by $B$ given in equation (\ref{beta}).

\subsection{Numerical Method}
As the shock emerges from the center of our model elliptical galaxy
and moves outwards, it will accelerate particles at each position it
passes through. Once the non-thermal particles are generated by the
shock, they will be frozen in the fluid element as we assume no
diffusion, and the spectrum will evolve according to equation
(\ref{timespec}). In practice, in our modeling, at every given time,
there will be non-thermal electrons within a certain radial range
with the new ones produced by the shock and the old ones moving out
with the fluid. Accordingly, we calculate the synchrotron emissivity
from equation (\ref{totalpower}) at each radius and then obtain the
total luminosity by integrating over the volume.

In practice, once the electrons are generated by a shock at certain
radius $r_0$ at time $t_0$, we follow the fluid element and compute
its position $r_t$ at time $t$. From the hydrodynamical code, we
know the velocity $v_r(t)$ and acceleration $a_r(t)$ of each fluid
element, so that the new position $r_t$ can be estimated as
\begin{eqnarray}
r_t=r_0+\sum_{t_0}^{t} \ \bigtriangleup r, \label{follow}
\end{eqnarray}
where the position increment $\bigtriangleup r$ in each time step
$\bigtriangleup t$ is calculated as
\begin{eqnarray}
\bigtriangleup r=v_r(t)\times\bigtriangleup
t+\frac{1}{2}a_r(t)\times\bigtriangleup t^2\ , \label{deltar}
\end{eqnarray}
For higher numerical accuracy, the quantities $v_r(t)$ and $a_r(t)$
of the fluid element at radius $r$ could be approximated by the
average velocity and acceleration of the two fluid elements at
radius $r$ and $r+dr$, which are given in the output from the code.

So, in our calculation, when a shock is found at time $t$, its
position is determined by the criterion given in equation
(\ref{shockcompression}). Then we follow the fluid element, that at
time $t$ was at the position of the shock 
according to equations (\ref{follow}) and (\ref{deltar}). After one
time step, we identify the new position of the shock and follow a
new fluid element. While following each
fluid element that passed the shock
, we calculate the value of $\Theta(t)$ given by equation
(\ref{Theta}). In this way, we know the energy spectrum of the
electrons from equation (\ref{timespec}) at different positions at
time $t$, and we finally  calculate the total synchrotron spectrum.
Note that our numerical method is different from the standard method
which solves the continuity equation for electrons in the 2-D phase
space at each time step (e.g., \citealt{longair1994}). Instead, in
our method, the energy spectrum at each time step is calculated
analytically according to equation (\ref{timespec}), and we only
need to follow one-dimensional the fluid element, which is
computationally more efficient.

\section{Results from one shock}
Now we give an example of the determination of the properties of the
synchrotron spectrum for a shock in model B3$_{02}$. In particular,
we focus on the last major burst in the example described in \S
\ref{codemodel}.

\subsection{Synchrotron Spectrum}
\begin{figure}
\includegraphics[angle=0.,scale=0.5]{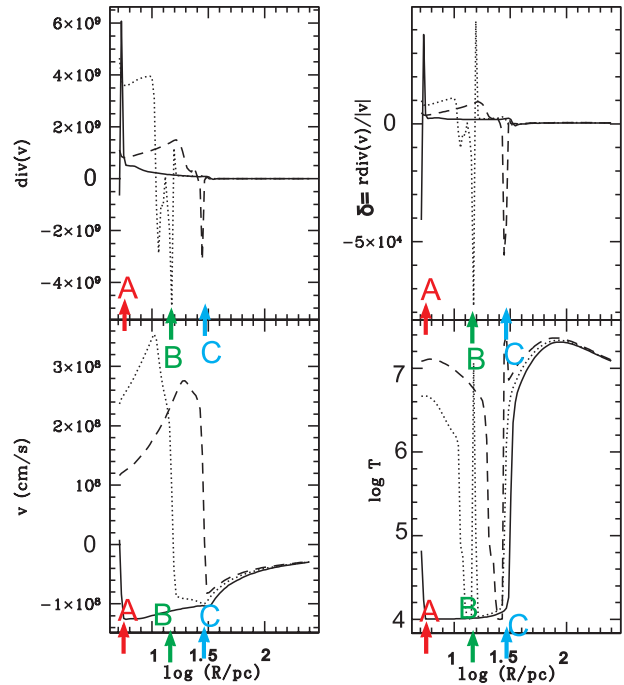}\\
\caption{The shock-locating method. In the two top panels, the
dimensional quantity $\rm{div}(\mathbf{v})$ (left) and the
dimensionless function $\delta$ (right, equation
\ref{shockcompression}) are shown for the same profiles shown in
Fig. \ref{examradiburst}. The solid, dotted and dashed lines are in
time order, separated by $0.004$ Myr. The solid line is at time
$\Delta t_1=0.235$ Myr after the time zero point. The three lines
also show the movement of the shock, which can be used to estimate
the shock velocity. There are three arrows and letters in the figure
to show the position of shock found at different times.
}\label{divshock}
\end{figure}

We define the time when the SMBH bolometric accretion luminosity
L$_{\rm{BH}}$ reaches 1 percent of Eddington luminosity
L$_{\rm{Edd}}$ as the zero time point, corresponding to 6.48 Gyr
in the code. Before the shock is formed and can be recognized by the
code, a cold shell is falling towards the center, which can be seen
clearly from Fig. \ref{examradi}. From Fig. \ref{examradiburst}, we
can see that at a later time $\Delta t=0.235$ Myr, the radial
velocity changes from negative to positive, indicating the emergence
of the nuclear wind and the shock. The result of equation
(\ref{shockcompression}) is shown in Fig. \ref{divshock}. At the
position of the shock, the parameter $\delta$ has a minimum value,
which confirms our criterion for finding the shock.

\begin{figure*}
\includegraphics[angle=0.,scale=0.5]{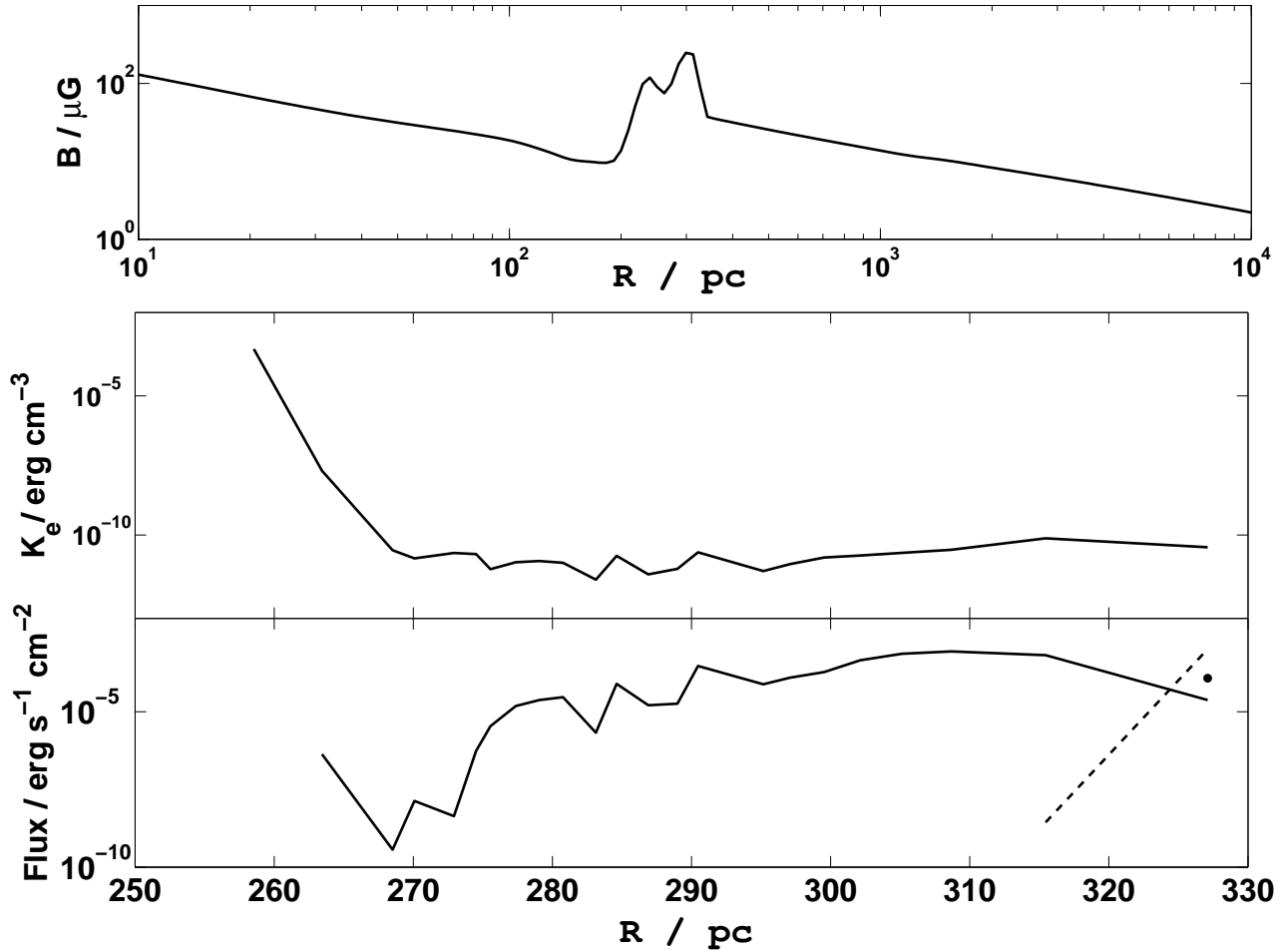}\\
\caption{Radial profile of the magnetic field, electron energy
density and synchrotron flux for three frequencies at time $\Delta
t=0.335$ Myr. At this time, the shock is located at about $327$ pc.
The upper panel shows the radial profile of magnetic field estimated
according to equation (\ref{beta}) at this time. The bump around
$300$ pc is due to the compression by the shock. The middle panel
shows the radial profile of electron energy density calculated
according to equation (\ref{elecenergy}). As we assume no diffusion,
non-thermal electrons only exist in the radial region $258$ ---
$327$ pc at this time. The lower panel shows the radial profile of
the flux for different frequencies: radio band $4.84$ GHz (solid
line), optical band $7.25\times 10^{14}$ Hz (dash line), X-ray band
$2.42\times 10^{17}$ Hz (the dot). The flux at radius $R$ is
estimated by $\nu J_{\nu}\times R$. Note that the higher the
frequency is, the closer the flux is to the shock. At this time, the
X-ray emission can only be found just after the shock, while the
optical emission can be found in two cells after the shock. For the
radio emission, it exists in a much more extended region.
}\label{radislprofile}
\end{figure*}

\begin{figure*}
\includegraphics[angle=0.,scale=0.6]{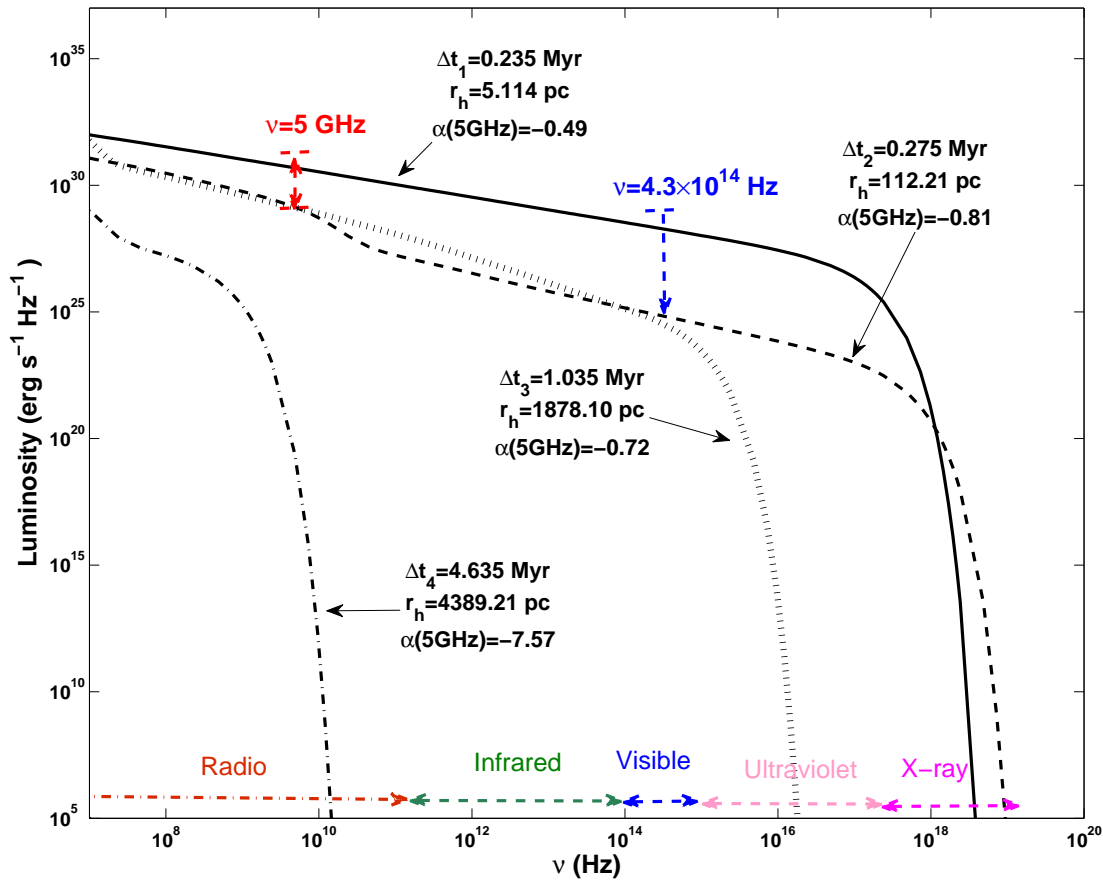}\\
\caption{Synchrotron emission spectra without absorption at four
different times. The solid, dash, dot and dash-dot lines are the
total synchrotron spectrum from the elliptical galaxy at the time
$\Delta t_1=0.235$ Myr, $\Delta t_2=0.275$ Myr, $\Delta t_3=1.035$
Myr and $\Delta t_4=4.635$ Myr respectively after the defined time
zero. The half radius $r_h$ with each line is the radius within
which half energy at the frequency $5$ GHz is emitted. $\alpha$ is
the spectral index defined in equation (\ref{spectralindex}). The
arrow labeled with `$\nu=5$ GHz' is the range of observed core radio
luminosity at $5$ GHz in FR II sources (\citealt{ZirbelBaum1995};
\citealt{Chiabergeetal2000}). The arrow labeled with `$\nu=4.3\times
10^{14}$ Hz' is the range of observed core optical luminosities at
$4.3\times 10^{14}$ Hz (wavelength $\lambda=7000 \dot{A}$) in FR II
sources by $HST$, which are believed to originate from non-thermal
emission \citep{Chiabergeetal2000}. At the bottom of this figure,
the names for different frequency bands are labeled.
}\label{diffspectrum}
\end{figure*}

\begin{figure}
\includegraphics[angle=0.,scale=0.28]{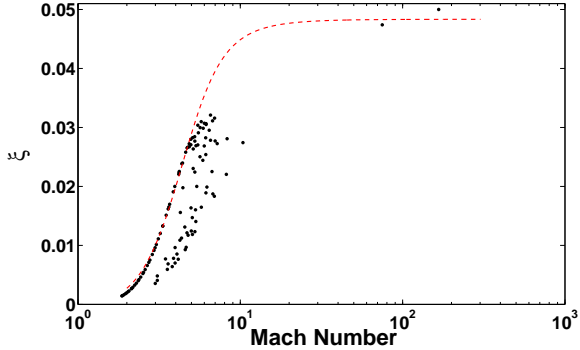}\\
\caption{Acceleration efficiency at different Mach numbers,
$\mathcal{M}$. The figure shows the ratio between the non-thermal
electron energy density accelerated by the shock to the downstream
thermal energy density just after the shock $\xi$, as computed from
the hydrodynamical code. For high $\mathcal{M}$, the ratio is almost
$0.05$, which occurs when the shock encounters the cold shell. For
small $\mathcal{M}$, the ratio can be around $0.001$. To the degree
that $\xi \ll 1$, the test particle theory adopted in this paper is
accurate. Roughly speaking, when $\mathcal{M}$ is larger than 10,
modification to the normalization (equation \ref{saturation}) is
important. The red dashed line is the theoretical result of the
model with an assumed constant downstream temperature $5\times10^7$
K, an assumed constant downstream  number density $10$ cm$^{-3}$,
and an assumed maximum electron energy $100$ TeV.  Even for the same
$\mathcal{M}$ the ratio $\xi$ can be different for different
temperature and number density.}\label{eratio}
\end{figure}

\begin{figure}
\includegraphics[height=6cm]{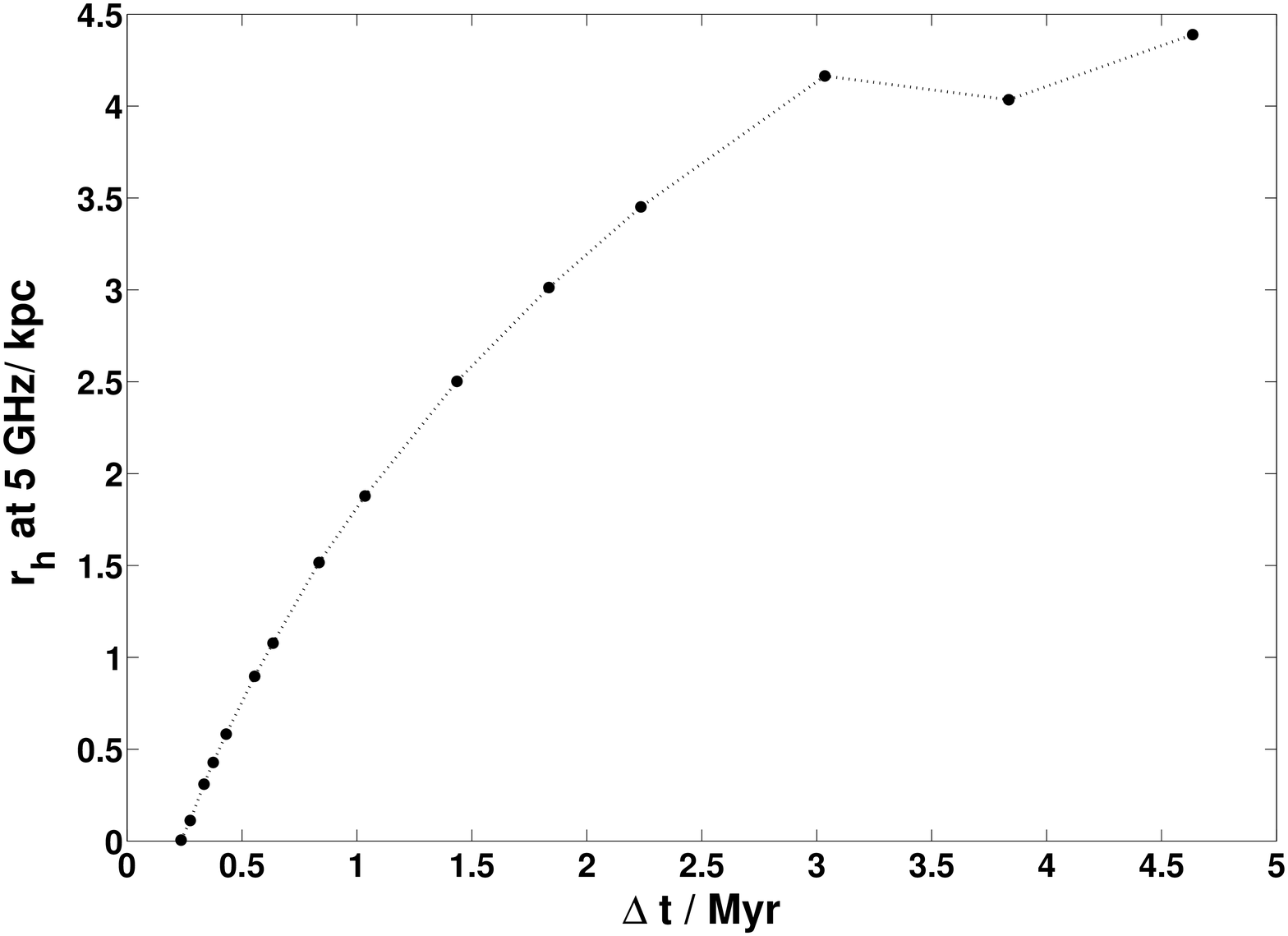}\\
\caption{Evolution of the synchrotron radio emission size at $5$
GHz. The size $r_h$ is defined to be the radius within which half
the energy at the frequency $5$ GHz is emitted. At around $3$ Myr,
the size decreases because at this time there is no contribution to
the $5$ GHz synchrotron radio emission from the new shock position
due to small magnetic field. Then the emission at this band is only
from the `old' electrons, which suffer large radiation and adiabatic
losses and move outwards. }\label{size}
\end{figure}

At each time we find the position of the shock and follow the shock
until it disappears. Meanwhile, we calculate the energy spectrum of
the non-thermal electrons accelerated by the shock with equation
(\ref{realspec}) and the spectral evolution according to equation
(\ref{timespec}). Then the synchrotron emissivity at each radius is
calculated from equation (\ref{totalpower}) and the total
synchrotron spectrum for our model elliptical galaxy is the sum of
the emission from different positions. So, the final spectrum is
actually not from a single population of power law electrons, but
from electrons at different times in their evolution. In Fig.
\ref{radislprofile} we show the radial profile of the magnetic
field, the non-thermal electron energy density and the flux for
different frequencies at the time $\Delta t=0.335$ Myr. The magnetic
field is estimated according to equation (\ref{beta}). At this time,
the shock is located at about $327$ pc and the bump of the magnetic
field is due to the compression at the shock. The non-thermal
electron energy density is estimated according to equation
(\ref{elecenergy}). As we assume no diffusion, non-thermal electrons
only exist in the radial region $258$ --- $327$ pc at this time. The
non-thermal electrons around $260$ pc come from the initial strong
shock (arrow A in Fig. \ref{divshock}). The flux at radius R is
estimated by $\nu J_{\nu}\times R$, and the fluxes of radio emission
(4.84 GHz), optical emission ($7.25\times 10^{14}$ GHz) and X-ray
emission ($2.42\times 10^{17}$ GHz) are labeled by solid, dashed and
dotted lines, respectively. Though the electron energy density
around $260$ pc is large, these are ``old" electrons that do not
contribute to the synchrotron emission in the three frequencies by
this time. The X-ray emission can only be found just behind the
shock, while the optical emission can be found within two cells of
the shock. However, the radio emission can extend over almost the
whole region passed by the shock.

In our model, the only free parameters needed to calculate the
synchrotron emission are $x_{inj}$, defined in equation
(\ref{minmomentum}), which is related to the minimum momentum of the
accelerated electrons, and the saturation parameter
$\xi_{max}=0.05$. In order to make the injection coefficient $\eta$
(equation \ref{eta}) to be about $10^{-4}$, which is a reasonable
value used in the models of SNRs (e.g.,
\citealt{Berezhkoellison1999}), we take the value\footnote{If
$x_{inj}$ is smaller, more electrons will be accelerated and the
emission will be stronger while the contrary is also true. However,
as electrons with momentum around $p_{min}$ do not contain most of
the energy for non-thermal electrons in strong shocks, the total
spectrum is not very sensitive to $x_{inj}$.} $x_{inj}=3.6$, the
same as the value taken in the model of \cite{Ensslinetal2007}. Note
that as we do not include the jet feedback in the hydro simulation
(see Paper I), the synchrotron emission calculated here is not the
total emission that would be observed from a real elliptical galaxy.
It can only be compared to elliptical galaxies without jets or to
the core emission in elliptical galaxies with extended jets. We take
snapshots at four different times during the evolution of the model.
The resulting spectra are shown in Fig.~\ref{diffspectrum}. The
change of spectral shape with time in this figure is due to the
radiative and adiabatic losses. In particular, the high frequency
part disappears at late times (due to a smaller magnetic field at
larger radii and
 losses). We define the
fiducial radius $r_h$ enclosing the volume within which half the
energy at the frequency $5$ GHz is emitted. The spectral index
$\alpha$ at a certain frequency is defined to be the index of the
power law approximation:
\begin{eqnarray}
J_{\nu}\propto\nu^{\alpha}\ .
\end{eqnarray}
In practice, we calculate the spectral index by the following
difference
\begin{eqnarray}
\alpha=\frac{\log J_{\nu_1}-\log J_{\nu_{2}}}{\log \nu_1-\log
\nu_2}\ , \label{spectralindex}
\end{eqnarray}
where the two frequencies are chosen to be $\nu_1=0.17$ GHz and
$\nu_2=5$ GHz, so that we can compare our calculation with
observations in the two bands. From Figure \ref{diffspectrum}, we
can see that the spectrum becomes steeper with time and the absolute
value of the spectral index becomes larger. The ratio $\xi$ of the
non-thermal electron energy density generated by the shock to the
downstream thermal energy density  will change for different Mach
numbers, as shown in Fig. \ref{eratio}. The non-thermal electron
energy density is calculated according to equation
(\ref{elecenergy}). Note that the maximum ratio is $\xi\sim0.05$,
which is nearly reached only for very high Mach numbers. For small
Mach numbers, the ratio is only about $0.001$. Roughly speaking,
when $\mathcal{M}$ is larger than 10, the saturation condition
(equation \ref{saturation}) becomes very important. Though we do not
consider protons here, they will also be accelerated by the shock.
As total cosmic ray energy density is estimated to be around $10$
percent of the shock's kinetic energy (e.g.,
\citealt{Aharonianetal2004}), the energy density ratio between the
protons and electrons will also change with Mach number. For large
Mach number, the energy density ratio between the non-thermal
protons and non-thermal electrons is smaller (say 10 or smaller)
while for small Mach number, the ratio is larger (for example 100).
The adopted test particle theory assumes that the post shock CR
pressure is small compared to the thermal gas pressure and we find
that this assumption to be valid during most of the time of
evolution except when the shock encounters the cold shell, where the
modified spectrum takes effect. To the first order of approximation,
we neglect the pressure of CRs, which is a first step to determine
whether a more advanced model is merited.

\begin{figure}
\includegraphics[height=6cm]{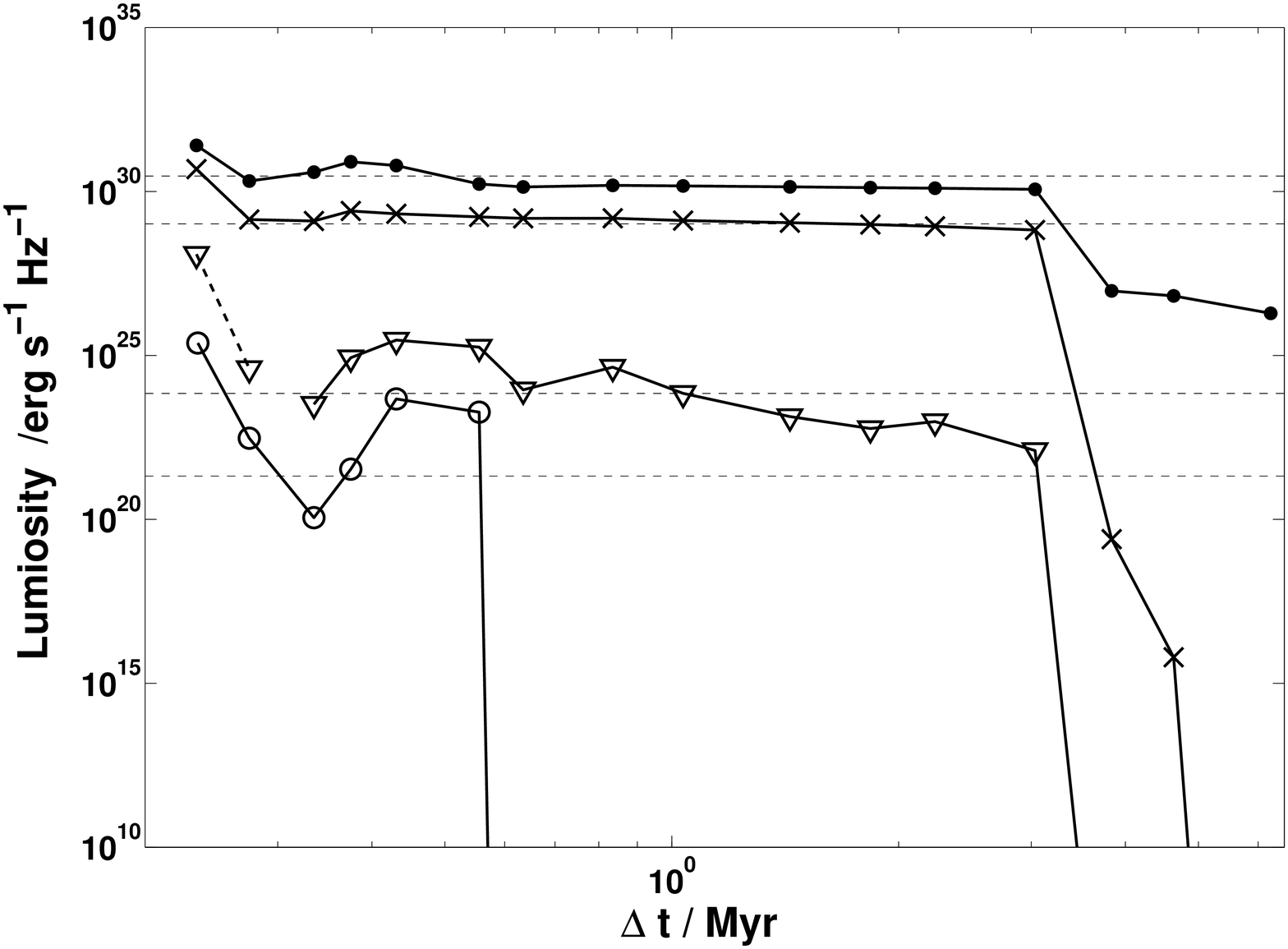}\\
\caption{Evolution of the synchrotron luminosity in 4 different
bands: $\nu=0.17\ \rm{GHz}$ (solid dots, radio), $\nu=4.84\
\rm{GHz}$ (crosses, radio), $\nu=7.25\times 10^5\ \rm{GHz}$
(triangle, optical, and the first two points in this line are
connected by a dash line because they will be absorbed by the dust),
$\nu=2.42\times 10^8\ \rm{GHz}$ (open circles, X-Ray).
 The larger the frequency is, the sooner radiation at that frequency
disappears. In the four horizontal dashed lines we show what would
be seen if $\nu L_{\nu}$ for each cited frequency band had the power
$5\times 10^{38}$ erg s$^{-1}$. }\label{frequencies}
\end{figure}

\subsection{Synchrotron Radio Emission}

While the shock moves outwards, it dissipates energy into the ISM
and  finally disappears. In this example, we can still see the shock
at the time $\Delta t=46$ Myr at the radius $32.7$ kpc, where a
second shock due to the next sub-burst emerges. But as we have shown
in Fig. \ref{diffspectrum}, the radio emission is already almost
unobservable at the time $\Delta t =4.635$ Myr. In Fig. \ref{size},
we show the evolution of the synchrotron emission size $r_h$ (at $5$
GHz, defined as the spatial radius enclosing half of the emitted
synchrotron luminosity at the given time). Generally, $r_h$
increases with time as the non-thermal electrons, which are frozen
in the fluid element, move outwards. However, when the shock
propagates into the regions with low magnetic field strength, there
will be almost no synchrotron radio emission at $5$ GHz and the
apparent size will decrease at that time. The size $r_h$ may change
from a very small initial size (5.1 pc in this example) to around
4.5 kpc. In Fig. \ref{frequencies}, we show the evolution of
synchrotron luminosity at 3 different frequencies. Note that the
spectrum shown in Fig. \ref{diffspectrum} is the emission spectrum
without allowance for absorption. However, for radio bands, the
optical depth due to synchrotron self-absorption at $0.1$ GHz
initially is only about $10^{-4}$ and smaller at later times. The
synchrotron self-absorption is thus only important for frequencies
smaller than $10^7$ Hz in this example.

\begin{figure*}
\includegraphics[angle=0,scale=0.5]{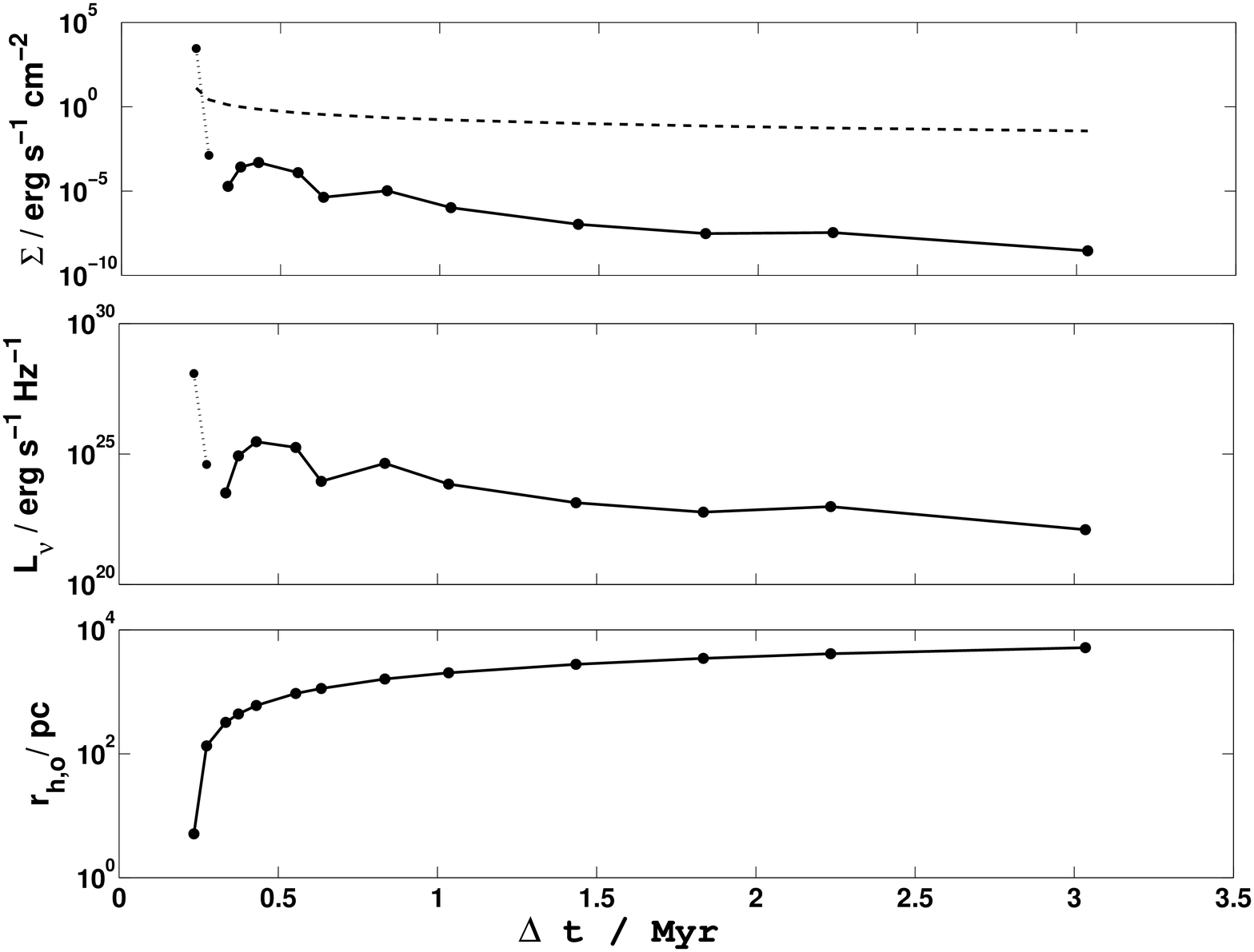}\\
\caption{ Properties of synchrotron (optical) emission in model
B3$_{02}$. This figure shows the time evolution of the flux
 $\Sigma$ (upper panel), luminosity $L_{\nu}$ (middle
panel) and size $r_{h,o}$ (lower panel) at the frequency
$7.25\times10^{14}$ Hz. The dashed line in the top panel is
$\Sigma_{\star}(r_{h,o}/R_e)$, i.e., the stellar flux of a de
Vaucouleurs profile with $\Sigma_{e}=22$  mag/arcsec$^2$ and
$R_e=6.9$ kpc. In the upper and middle panels, the first two points
are connected by dotted lines because the optical emission will be
absorbed by dust at that time. Note that synchrotron emission lasts
for 2.8 Myr and then drops to zero very quickly. }\label{optical}
\end{figure*}

\subsection{Synchrotron Emission in Other Bands}

For optical bands, the main absorption is due to dust. The optical
depth due to dust absorption is calculated in the code (Paper I).
The optical emission is absorbed by dust for $\Delta t\le 0.33$ Myr
(within 300 pc in this example), but the absorption is not important
after that time. A fiducial estimate of the expected mean flux
$\Sigma_o$ of the optical synchrotron emission is obtained as
$\Sigma_o\equiv L_o/(2\pi r_{h,o}^2)$, where $L_o$ is the total
synchrotron luminosity in the optical band ($7.25\times 10^{14}$
Hz), and where we have neglected the difference between the radius
within which half of luminosity is emitted $r_{h,o}$ and the
projected surface effective radius. We compare $\Sigma_o$ with the
optical flux of our galaxy model, to see whether the synchrotron
optical emission can be observed or not. In order to estimate the
stellar flux of the galaxy model, we adopt as a fiducial value for
the surface brightness at the effective radius $\Sigma_{e}=22$
mag/arcsec$^2$, and for the effective radius $R_e=6.9$ kpc.
At each time, we then compare the value of $\Sigma_o$ with the
stellar flux profile obtained from the de Vaucouleurs profile, i.e.,
with $\Sigma_{\star}(r_{h,o}/R_e)$. The result is shown in Fig.
\ref{optical}. This figure shows that the synchrotron optical
emission is stronger than the star light only at early times. Later
on, it is too faint compared with the star light. However, the
optical emission will be absorbed by the dust in this example at the
early time. So we expect to observe synchrotron optical emission at
the inner part of those elliptical galaxies with little dust during
a short time in the quasar phase. This result is just from one shock
at the first sub-burst of the last major burst in Fig.
\ref{examtimeburst}. However, we stress that within each burst
several shocks appear from the sub-bursts in the galaxy almost
simultaneously, which will give stronger emission and a higher
probability of observing the synchrotron optical emission. The study
of the cumulative effect of several co-existing shocks is beyond the
reach of the present investigation, and it is postponed to a future
work. In particular, it will be important also to consider the more
realistic case of multiple shocks created by the so-called
``time-dependent nuclear wind treatment" which includes the time for
the wind to reach different radii (see Paper I , see also \S
\ref{discussion}). Right now, we have just checked that, for a
single shock, the time-dependent and time-independent wind
(neglecting the time needed for the wind to reach different radii)
models do produce very similar results.

\begin{figure*}
\includegraphics[angle=0,scale=0.5]{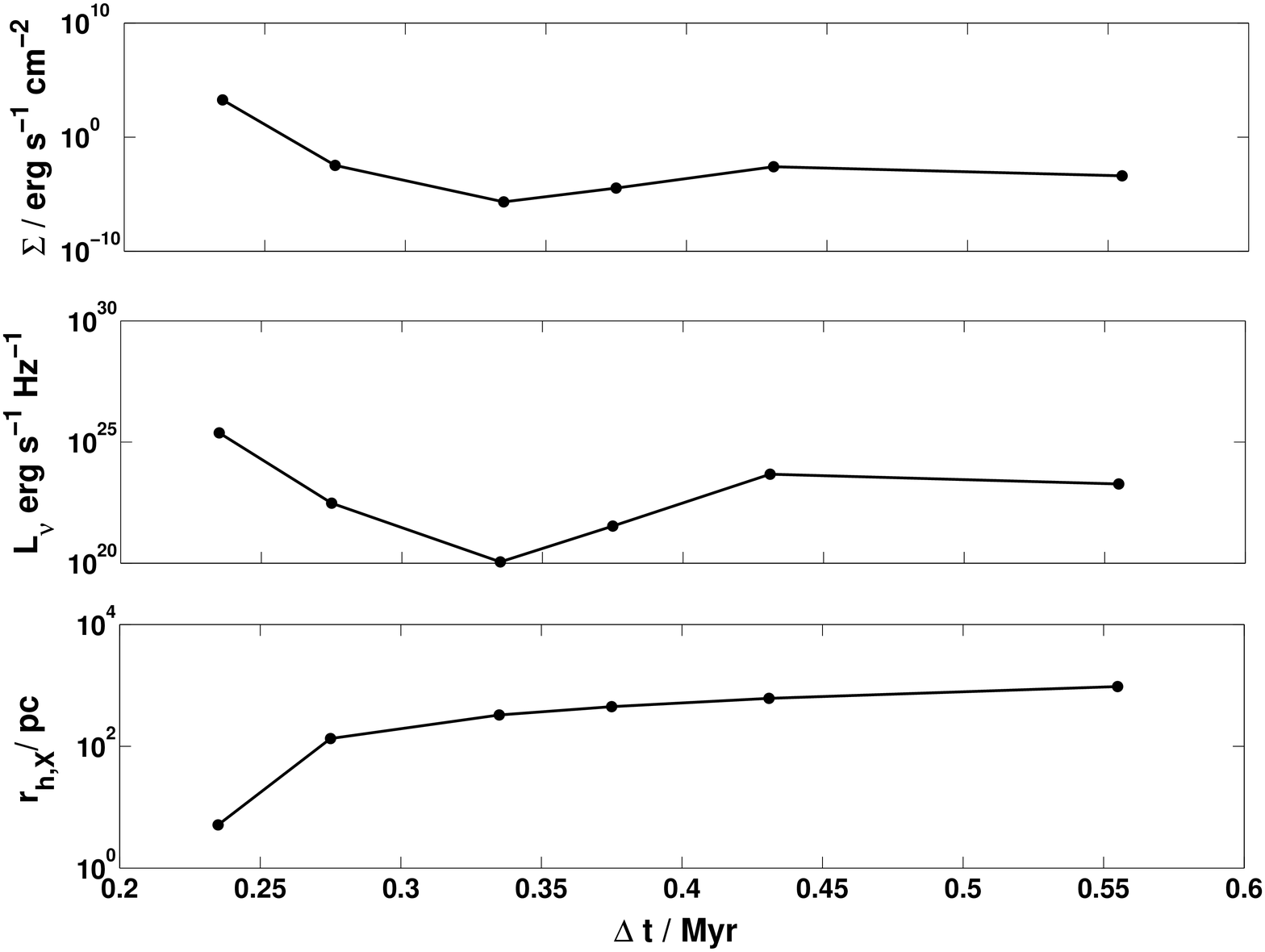}\\
\caption{Properties of the synchrotron (soft X-ray) emission in
model B3$_{02}$. This figure shows the time evolution of the flux
 $\Sigma$ (upper panel), luminosity $L_{\nu}$ (middle
panel) and size $r_{h,x}$ (lower panel) at the frequency
$2.41\times10^{17}$ Hz at different times. The half radius in the
X-ray band is almost the same as the radius of the shock, which
means the synchrotron X-ray emission only exists within a narrow
region around the shock. }\label{xray}
\end{figure*}

Fig. \ref{diffspectrum} shows that there is also synchrotron X-ray
emission for the duration of $\sim10^4$ yr to $\sim10^5$ yr at the
very early time of the burst. At the time $\Delta t=0.235$ Myr, at
the frequency $2.4\times10^{17}$ Hz (equivalent 1 keV), the
luminosity is $5.76\times10^{42}$ erg s$^{-1}$ while at the
frequency $1.2\times10^{18}$ Hz (equivalent 5 keV), the luminosity
is $1.2\times10^{38}$ erg s$^{-1}$. The synchrotron X-ray emission
lasts for only 0.32 Myr and drops to zero rapidly after that time.
Typically the luminosity of the thermal X-ray emission in the band
$0.5-2.0$ keV from isolated elliptical galaxies varies from
$10^{39}$ to $10^{41}$ $\rm{erg}\ \rm{s}^{-1}$ (e.g.,
\citealt{Memolaetal2009}). So the synchrotron X-ray emission are
likely to  be observed only in these ``on" state galaxies, which are
only a small fraction of all galaxies. Because the spectrum of
bremsstrahlung emission (energy density per Hz) is almost flat, we
should expect the luminosity of bremsstrahlung optical emission to
be around $10^{35}$ to $10^{37}$ $\rm{erg}\ \rm{s}^{-1}$ based on
the bremsstrahlung X-ray emission. This is actually smaller than the
synchrotron optical emission here.

As synchrotron X-ray emission is also observed from SNRs, we can
scale down the flux of the X-rays calculated here to a typical flux
of X-rays from SNRs. If synchrotron emission from SNRs and
elliptical galaxies is from the same mechanism and we keep the ratio
of synchrotron optical emission and synchrotron X-ray emission
calculated here, we can estimate the flux of synchrotron optical
emission from SNRs and see if it can be observed. Because the
synchrotron spectrum basically becomes steeper with time, we can use
the ratio at the initial time to estimate the lower bound of the
synchrotron optical emission from SNRs. The synchrotron X-ray flux
observed from SNR G330.2+1.0 is about $10^{-11}$ erg s$^{-1}$
cm$^{-2}$ \citep{Parketal2009}. In our calculation, at the time
$\Delta t=0.235$ Myr (see the solid line in Fig.
\ref{diffspectrum}), the X-ray flux (at 1 keV) is about $1.85\times
10^{3}$ erg s$^{-1}$ cm$^{-2}$ while the optical flux is about
$2.83\times 10^{3}$ erg s$^{-1}$ cm$^{-2}$. Then we give a lower
bound of synchrotron optical emission from SNRs to be
$1.53\times10^{-11}$ erg s$^{-1}$ cm$^{-2}$, corresponding to
$44.87$ mag/arc$^2$ without absorption. Actually, in SNR G332.5-5.6,
line emission with a flux as faint as $10^{-12}$ erg s$^{-1}$
cm$^{-2}$ has been observed \citep{Stuparetal2007}. It seems that
the synchrotron optical emission from SNRs should be observable by
the appropriate technique.

\section{Discussion}
\label{discussion}

\subsection{Observational Consequences of Our Model}

Our model has observational predictions, some of which are
consistent with current radio observations and some of which may be
feasible $Chandra$ source detections. First, as shown in Fig.
\ref{diffspectrum}, the synchrotron emission at 5 GHz and
$4.3\times10^{14}$ Hz falls in the range of observed values in FR II
sources, which means that our mechanism should not be neglected.
Besides, recent observations at 1.4 GHz find some early-type core
galaxies hosting radio-loud AGN of extremely low radio power
(\citealt{BaldiCapetti2009}), which can be interpreted as radio
emission via the mechanism described here as it would be too weak to
be produced by typical jets. Second, as shown in Fig. \ref{size},
the shocks produced in the simulations easily reach the kpc scale
before they die away gradually. The synchrotron radio emission can
also go to this large scale (though it is very weak). Actually,
kpc-scale outflows or winds have been observed in high red-shift
quasars (e.g., \citealt{dekooletal2001}; \citealt{Nesvadbaetal2008};
\citealt{Nesvadba2009}; \citealt{Cattaneoetal2009}), which provides
strong support for our scenario based on the computed winds in the
AGN simulations. In particular, a recent observation
(\citealt{Alexanderetal2009}) finds a galaxy wide outflow at
redshift $\sim$ 2, which is believed to be a wind radiatively driven
by the AGN and/or supernovae winds rather than by jets. Note that
some observed X-shaped radio galaxies (e.g.,
\citealt{LeahyWilliams1984}; \citealt{Kraftetal2005};
\citealt{saripallietal2008}) provide significant support for the
idea that the observed radio emission arises from shocks caused by
feedback from the SMBHs, as the broad line winds drive shocks into
the ambient gas with a similar shape (e.g., \citealt{DrewProga2000};
\citealt{Dorodnitsynetal2008}).

Third, the duration of synchrotron emission at a certain frequency
will decrease with increasing frequency. For a single shock,
synchrotron X-ray emission can only exist behind the shock, where
the non-thermal electrons are just created. The synchrotron optical
emission can extend to a larger region while the synchrotron radio
emission can almost extend to the whole region where the shock has
passed, which can be seen from Fig. \ref{radislprofile}. So, in
principle, at good enough resolution we should see the synchrotron
emission in different bands at different spatial positions.

Though synchrotron emission from our mechanism and jets is similar
to a certain extent, there are some differences that can help
distinguish the two mechanisms. In the following, we list four
possible observational tests. First, we expect the magnetic field in
the shocked material to be more randomized than the magnetic field
in the jets. So, the polarization of the emission from our mechanism
will be weaker than the polarization of the emission from the jets,
and polarization observations can help to disentangle the two
mechanisms. Second, as the shock (and the synchrotron emission) is
produced after each AGN outburst, emission from our mechanism will
only exist in a certain fraction of elliptical galaxies,
proportional to the duty cycle. In practice, in a survey of
elliptical galaxies, where the emission from jets gives the
background level of core radio emission, the number of galaxies with
extra radio emission should be comparable to the duty cycle (see
Paper I for a discussion of duty cycle of the models). Third, as the
velocity of the jet is typically much larger than the velocity of
the shock, then if we can directly measure the velocity of the
outflow, it will be easy to tell which process dominates. Finally,
the variability time scale of emission from jets is shorter than
that of our mechanism. The variability time scale can also help us
tell the origin of the core radio emission.

\subsection{Possible Future Improvements of the Model}

There are still some significant uncertainties in our model and
further improvements are needed. First, the hydrodynamical code does
not include the magnetic field, a fundamental ingredient for the
accurate computation of synchrotron emission. In this paper, we only
assumed a ratio between the magnetic pressure and thermal pressure
to estimate the preshock magnetic field in elliptical galaxies. A
better method is needed to give the distribution of magnetic field
and its evolution with time.

In fact, the magnetic field will be amplified near the shock due to
back reaction of cosmic rays, and we roughly estimate that the
amplified magnetic pressure should not exceed
 about $10\%$ of the postshock thermal pressure. A more careful
calculation of this effect (e.g., \citealt{Bell2004};
\citealt{RiquelmeSpitkovsky2009}; \citealt{Ohiraetal2009};
\citealt{Zirak08}) would be preferable. Second, we use the
test-particle theory to calculate the energy spectrum of the
non-thermal electrons and neglect the modification of the shock
structure due to the cosmic rays. However, as is already known in
SNRs, non-linear theory is needed to calculate the spectrum of the
non-thermal electrons for strong shocks, especially at high energies
(e.g., \citealt{Berezhkoellison1999}; \citealt{Lazendicetal2004};
\citealt{Leeetal2008}; \citealt{TreumannJaroschek2008};
\citealt{Reynolds2008}).
So in order to calculate the synchrotron emission more accurately,
we will need to include the diffusive shock acceleration mechanism
into the hydrodynamical code. This greatly adds to the complexity of
the modeling. Third, as anticipated in \S \ref{codemodel}, all the
results presented here are obtained using a 1-D code. Of course,
this is quite a strong approximation, which precludes the
quantitative analysis of several phenomena that almost certainly
take place in real galaxies. One of the most important in the
present context is the possible cold shell fragmentation due to
Rayleigh-Taylor instability following central outburst. While we
could not exclude this possibility, we note that several properties
of our 1-D models (such as the duty-cycle, mass accreted on the
central SMBH, star formation, X-ray luminosity) agree well with the
observations, so that we are confident that the basic physics of
feedback is actually captured by our simulations. Note also that on
the kpc scale, the central outbursts, even if initially anisotropic,
will necessarily become rounder and rounder as they propagate, and
therefore better and better approximated by our code.

The calculations here  are based on the so-called ``time-independent
nuclear wind model" (with mechanical as well radiation feedback). In
this simplified approach, which saves much simulation time, the
propagation velocity of the wind along the numerical grid (Paper I,
eq. 29) is set to be infinity. Therefore, the nuclear wind mass,
momentum and kinetic energy are instantaneously discharged over the
whole computational grid. The more realistic model is the
time-dependent nuclear wind model, which will be more accurate,
especially at the inner part of the galaxy at early times (\S 2.3 of
Paper I). More accurate calculations of the synchrotron optical and
X-ray emission also require the time-dependent wind model. Finally,
effects of multiple shocks on the emission from the inner galaxy and
on the reacceleration of preexisting nonthermal particles have yet
to be included.

These caveats aside, the presented results show that central
outbursts of mechanical and radiative energy consequent to AGN
flaring should produce shock accelerated electrons capable of
producing the non-thermal emission actually observed from the cores
of some elliptical galaxies subsequent and consequent to AGN
outbursts.

\section{Conclusions}
In this paper, we have explored the observational consequences of a
new mechanism for producing the synchrotron emission seen in
elliptical galaxies. This idea is based on particle acceleration
subsequent to bursts of energy from the central black holes. The
properties of the ISM are computed from hydrodynamical simulations
including radiative and mechanical feedback consequent to an AGN
outburst. Due to this feedback, a wind is formed during the quasar
phases, and a shock is driven into the elliptical galaxy. Given
standard physics, the shock, during its propagation into the galaxy,
will accelerate electrons and protons via the diffusive shock
acceleration mechanism. Synchrotron emission arises from these
non-thermal electrons. In this paper, we have focused on the
computation of the energy spectrum of the integrated synchrotron
radiation. We include radiation loss and adiabatic loss during the
evolution of these non-thermal electrons and we show the change of
the synchrotron spectrum with time.

In general, we have found that the synchrotron radio emission from
one shock can last about $4.7$ Myr at $5$ GHz, while a shock can
exist for $46$ Myr. During the evolution of the elliptical galaxy,
there are several major bursts and several sub-bursts for each major
burst. Roughly speaking, the shock can exist during the whole burst
phase of the galaxy while the radio emission (around $5$ GHz) exists
for 10 percent of the burst phase. Besides, our mechanism also gives
marginally detectable synchrotron optical emission during the 3 Myr
after a burst. The duration of synchrotron optical emission is about
$50\%$ of the radio emission at $\sim0.17$ GHz while the duration of
the synchrotron X-ray emission is approximately $5\%$ of the radio
synchrotron emission at $\sim0.17$ GHz. The half radius of the
synchrotron X-ray emission is almost the same as the radius of the
shock, which means that the synchrotron X-ray emission only exists
as a `ring' around the shock while the half radii of the synchrotron
radio emission and synchrotron optical emission are smaller than the
shock radius, which indicates the synchrotron radio emission and
synchrotron optical emission are more diffuse. Actually, synchrotron
optical emission from the cores of elliptical galaxies has been
claimed to be observed by $Hubble\ Space\ Telescope\ (HST)$ with a
typical luminosity $\sim10^{25}-10^{29}$ erg s$^{-1}$ Hz$^{-1}$
(e.g., \citealt{Chiabergeetal2000}). The detection of synchrotron
X-ray emission from elliptical galaxy cores
is on the edge of currently feasible $Chandra$ observations (e.g.,
\citealt{Tanabaum2001}, and private communication).

One point which needs to be emphasized is that the mechanism for
synchrotron radio emission discussed here is different from that
produced by jets. Current understanding of the AGN phenomena divides
the activity into two phases: the `optical mode', immediately after
the central outburst, and `radio mode' at low Eddington ratios
between major outbursts, when powerful jets are emitted. In
particular, here we have discussed the particle acceleration during
the optical mode, which should exist for each quasar event
associated with the outburst phase. Though the synchrotron radio
emission from large scale jets may be stronger than the presented
mechanism, our mechanism is also important for the synchrotron
emission at the central part of elliptical galaxies and it could
also be important for the evolution of the galaxy since the energy
is released within the gaseous envelope of the galaxy. Definite
observation of non-thermal emission from AGN hosting galaxy cores
would strengthen the case for AGN feedback. Though there may be
unresolved jets in the galaxy cores as pointed out by some observers
(e.g., Ho 2008 and references therein), our mechanism contributes
comparable emission to these small jets, if not larger.

\section*{ACKNOWLEDGMENT}
Y.-F. Jiang thanks Carlos Badeness and John Hughes for helpful
discussions on diffusive shock acceleration mechanism. We also thank
Roberto Fanti, Luis Ho, Jenny Greene and Aristotle Socrates for
useful discussions on cosmic rays and observational results, and Yue
Shen, Min-Su Shin and Lorenzo Sironi for useful discussions. We also
thank two anonymous referees for comments that significantly
improved the paper. Y.-F. Jiang thanks Princeton University for
financial support. A.S. acknowledges support from NSF grant
AST-0807381.

\end{document}